%% file: gieseler18a.tex
\documentclass[
prl,onecolumn
]{revtex4-1}

\usepackage{graphicx}
\usepackage{amsmath}
\usepackage[mathscr]{eucal}

\usepackage{upgreek}

\include{commands}

\begin{document}
\title{Levitated Nanoparticles for Microscopic Thermodynamics - A Review}

\author{Jan Gieseler}
 \affiliation{Department of Physics, Harvard University, 17 Oxford Street, Cambridge, MA 02138, USA}
\author{James Millen}
 \affiliation{Faculty of Physics, University of Vienna, Boltzmanngasse 5, Vienna 1090, Austria}
 \affiliation{Department of Physics, Kings College London, Strand, London WC2R 2LS, UK}

\begin{abstract}
In this article, we review the current state of the art in using levitated nanoparticles to answer questions related to thermodynamics and non-equilibrium physics. We begin in Section 2 with a summary of the relevant deterministic and stochastic forces, which determine the particle dynamics and allow for control of the particle. In Section 3 we give a brief review of the stochastic (i.e. Brownian) motion of levitated particles, since the Brownian particle is fundamental to the theory of stochastic thermodynamics. Then we discuss the stability of the trapped particles and the related Kramers escape problem in Section 4. After that, we introduce effective potentials for the energy in Section 5. These potentials are useful to describe the dynamics of levitated nanoparticles in a time-modulated trap, where the particle is driven far away from equilibrium. In Section 6 we discuss the dynamics of relaxation towards equilibrium, before we review the work on fluctuation theorems in Section 7. Fluctuation theorems are a powerful generalization of the well known thermodynamic inequalities for systems far from equilibrium. Finally, Section 8 discusses the potential of constructing new kinds of heat engines based on nanoparticles levitated in high vacuum.
\end{abstract}

\maketitle


\section{Introduction}

In 1827, botanist Robert Brown noted the erratic movement of tiny particles emitted from pollen grains in a liquid \cite{Brown:1828ut}. This seemingly unspectacular observation would play a critical role in the development of the atomistic theory of matter. However, only in 1905 did Albert Einstein's theoretical analysis \cite{Einstein:1905gw} of Brown's observation provide crucial evidence for the existence of atoms. Einstein surmised that the random motion of the suspended particles is a consequence of the thermal motion of surrounding fluid molecules.
Ever since the \emph{{Brownian Particle}} has been essential in the development of our theoretical understanding of stochastic processes in different fields, ranging from the sciences (chemistry, biology and physics) to economics (e.g., finance).

Einstein further concluded that ``{\emph{the velocity and direction of motion of the particle will be already very greatly altered in an extraordinarily short time, and, indeed, in a totally irregular manner}}'' and that ``\emph{it is therefore impossible---at least for ultramicroscopic particles---to ascertain the instantaneous velocity by~observation}''.

This changed with the advent of optical tweezers, now a workhorse for studying thermodynamics and non-equilibrium physics of small systems.
The pioneering experiments on optical forces were carried out by Ashkin and Dziedzic, where they used optical forces to counteract gravity and thereby suspend micrometer-scale spheres in vacuum \cite{Ashkin:1970mb, Ashkin:1971dd, Ashkin:1976bp}.
Later Ashkin et al. demostrated the stable 3D optical trapping of micron-scale particles purely by light \cite{Ashkin1986}. Since then there has been an explosion of research using \emph{{optical tweezers}}, to the point that they are an off-the-shelf tool for physical and biological scientists \cite{JonesBook, Spesyvtseva2016}. In this system, it is possible to control and track the motion of mesoscopic objects with astounding precision.

While most of the research on optical tweezers has focused on trapping and manipulating particles in suspension, there has been a renewed interest in optical trapping in high vacuum \cite{Li2010, Gieseler:2012bi, Kiesel:2013bp, Millen2015}, motivated by the possibility to enter the quantum regime \cite{RomeroIsart2010, Chang2010, Barker:2010bw}, which has lead to the development of the field of levitated optomechanics \cite{Yin:2013bj, Vamivakas:2016tp}.
Even though there has been tremendous progress towards entering the quantum regime, with residual occupations of tens of phonons \cite{Jain2016a}, achieving the ground state has been elusive.
However, the exquisite control achieved in these experiments does not only bring us closer to the quantum regime, it also opens up a wide range of exciting new experiments in the classical domain.
In particular, they allow the study of Brownian motion of a single well-isolated particle with high temporal and spatial resolution and controllable coupling to the environment, thereby rebutting Einstein's original statement \cite{Li2010, Li2013} 
and providing new insights into microscale thermodynamic processes in the underdamped regime.
Specifically, thermodynamic processes of a single particle are stochastic, that is quantities such as energy, work and entropy are fluctuating quantities where the fluctuations are of similar magnitude or even larger than the mean. This has profound implications in the operation and fundamental limitations of microscopic machines.

Here we review the current state of the art in using levitated nanoparticles to answer questions related to thermodynamics and non-equilibrium physics.
We begin in Section 2 with a summary of the relevant deterministic and stochastic forces, which determine the particle dynamics and allow for control of the particle. In Section 3 we give a brief review of the stochastic (i.e. Brownian) motion of levitated particles, since the Brownian particle is fundamental to the theory of stochastic thermodynamics.
Then~we discuss the stability of the trapped particles and the related Kramers escape problem in Section~4.
After that, we introduce effective potentials for the energy in Section 5. These potentials are useful to describe the dynamics of levitated nanoparticles in a time-modulated trap, where the particle is driven far away from equilibrium.
In Section 6 we discuss the dynamics of relaxation towards equilibrium, before we review the work on fluctuation theorems in Section 7. Fluctuation theorems are a powerful generalization of the well known thermodynamic inequalities for systems far from equilibrium.
Finally, Section 8 discusses the potential of constructing new kinds of heat engines based on nanoparticles levitated in high vacuum.



\section{Forces and Potentials \label{sec:Forces}}
\subsection{Deterministic Forces}
\begin{figure}[h]
\begin{center}
  \includegraphics[width=\textwidth]{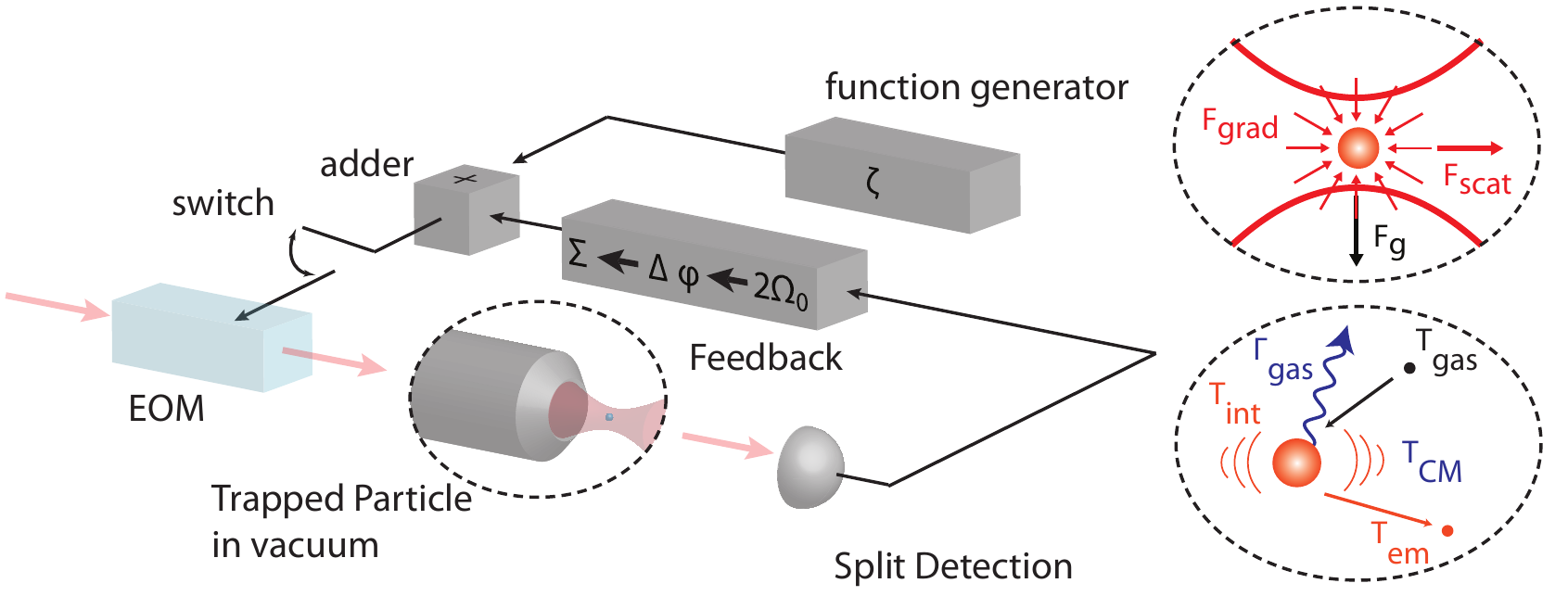}
  \caption{Schematic of optical levitation setup.
A nanoparticle is trapped by a tightly focused laser beam. The translational degrees of freedom of the nanoparticle are measured with photodetectors and the center-of-mass motion is cooled by parametric feedback. In addition to feedback, external modulation allows excitation of the particle to drive it far from equilibrium.
The top inset highlights the dominant forces in a typical optical levitation experiment, which are the optical gradient and scattering forces and gravity. The bottom inset shows the temperatures involved in a collision with a heated sphere: the sphere's centre-of-mass temperature ($\T{CM}$) and surface temperature ($\T{int}$), and the temperatures of the impinging gas particles ($\T{gas}$) and emerging gas particles ($\T{em}$) with $\T{gas}\leq \T{CM}\leq \T{em}\leq \T{int}$. The~collision with the air molecules leads to damping $\g{gas}$, which depends on the pressure.
  Main~figure taken from \cite{Gieseler:2014wt} with permission from Physical Review Letters. Inset adapted from \cite{Millen2014} with permission from Nature Nanotechnology.
\label{fig:Schematic}}
\end{center}
\end{figure}

Most experiments with particles levitated in vacuum use optical forces to create a stable trap (c.f. Fig~\ref{fig:Schematic}). This~gives a great deal of flexibility since optical fields can be controlled very well in intensity, position and time, allowing the creation of almost arbitrary force fields.
However, since light absorption heats~\cite{Hebestreit2018} and potentially destroys the particle, experiments have been limited to low absorption materials like Silica and Silicon.
In addition to the optical forces, the particle is subject to gravity {$\F{g} = m \mathbf{g}$}, electric forces $\F{e} = q\mathbf{E}$ if the particle carries a charge $q$ and magnetic forces $\F{mag} = \nabla(\boldsymbol{\mu}\cdot\mathbf{B})$ if the particle has a magnetic dipole moment  $\boldsymbol{\mu}$. 

Particles typically have a radius of $a\sim$ 100 $ \rm nm$ but can also be much bigger \cite{Li2010, Ashkin:1971dd}. When the radius is much smaller than the wavelength $\lambdaopt$ of the trapping laser $a \kopt \ll 1$, where $\kopt = 2\pi/\lambdaopt$, the particle can be treated as a dipole in the Rayleigh approximation. The polarizability for a particle with volume $V$ is thereby given by
\begin{equation}
\label{eqn:polarizability}
\polar{0} = \eo V \chitensor,
\end{equation}
where the total susceptibility of the particle  $\chitensor = \chitensor_e\left(1+\depol \chitensor_e\right)^{-1}$ depends on the material via the material susceptibility $\chitensor_e$ and on its geometry through the depolarization tensor $\depol$, which in general are both rank-2 tensors. However, for isotropic materials, the material susceptibility simplifies to a~scalar $\chi_e$ and similarly for a sphere the depolarization tensor is isotropic and simplifies to a scalar $N = 1/3$. Thus,~for a sphere we recover the Clausius-Mossotti relation $\chi = 3(\epsilon_p-1)/(\epsilon_p+2)$, where~we use $\epsilon_p= 1+\chi_e$.
For a particle with a uniaxial anisotropy, the susceptibility $\chitensor = \diag(\chi_\|, \chi_\perp, \chi_\perp)$, has a component $\chi_\parallel$ parallel and a component $\chi_\perp$ perpendicular to the symmetry~axis. For example, the depolarization tensor of a cylinder is $\depol = \diag(0, 1/2,1/2)$ in the frame of the cylinder, where the cylinder axis is along the $x$-axis. Consequently, $\chi_{\|} = \er - 1$, $\chi_{\perp} = 2(\er - 1)/(\er +1)$ for a cylinder with isotropic $\er$.
This means the maximal polarizability of a cylinder is $(\er+2)/3$ times higher than for a sphere of the equivalent volume, i.e. a factor of 2 for silica, and a factor of 4.6 for silicon.

In general, one has to consider the total field to calculate the optical forces \cite{LukasNovotny:ti}. The total field is the sum of the incident and the scattered field and is the self consistent solution to Maxwell's equations. For arbitrary shaped particles, the total field has to be calculated with numerical methods. 
However, for a spherical particle, the modified polarizability
\begin{equation}\label{eq:polarizability}
  \polar{} = \polar{0}\left(1-i\frac{\kopt^3\alpha_0}{6\pi\epsilon_0}\right)^{-1},
\end{equation}
accounts for the radiation reaction of the particle to its own scattered field, such that the induced polarization due to a field $\mathbf{E}_0$ is $\mathbf{P} = \alpha \mathbf{E}_0$. We introduce $\alpha'$ and $\alpha''$ to refer to the real and imaginary part of the polarizability, respectively.

Knowing the polarizability, we can calculate the optical force for sub-wavelength particles in the Rayleigh approximation. The optical force has conservative and non-conservative contributions \cite{Albaladejo:2009jq}
\begin{equation}\label{eq:Fopt}
  \F{opt} = 
\alpha'\nabla I_0/4 +\totcross\left[
\mathbf{S}/c+
 c \nabla\times \mathbf{L}\right],
\end{equation}
where $\totcross =\alpha''\kopt/\epsilon_0$, $\wopt$ is the optical frequency, and $\mathbf{L} = -i\epsilon_0\left<\mathbf{E}\times\mathbf{E}^*\right>\left/4\wopt\right.$, $\langle\dots\rangle$ representing a~time average. The total cross-section $\totcross$ is the sum of the absorption and scattering cross-sections. The first term is a conservative force $\F{grad} = \alpha'\nabla I_0/4$. It pulls particles with a high refractive index relative to their surroundings toward the region of maximum light intensity.
In optical tweezers, this is the focal volume of the light beam.

The non-conservative scattering force $\F{scat} = 
\totcross\left[
\mathbf{S}/c+
 c \nabla\times \mathbf{L}\right]$ has two contributions: the~radiation pressure term, which is proportional to the time averaged Poynting vector $\mathbf{S}=\left<\mathbf{E}\times\mathbf{H}^*\right>$, $\mathbf{H}$ being the magnetic field and a curl force associated to the non-uniform distribution of the time averaged spin density of the light field. The curl force is zero for a plane wave but can be significant for a tightly focused beam in optical tweezers.
However, since $\alpha''/\alpha'\propto a^3$, the non-conservative forces vanish for small particles and we will neglect them in the following discussion.

\subsubsection{Optical Potential}
The conservative force in Eqn.~\eqref{eq:Fopt} can be expressed as the gradient of a potential $\F{opt}\approx \F{grad} = -\nabla U_{\rm opt}$. 
At the bottom of the potential, the centre-of-mass motion is harmonic, with~frequencies

\begin{eqnarray}\label{eqn:frequencies_translational}
\w_{q} &= &2 \sqrt{\frac{\chi}{c\pi\rho }}\frac{\sqrt{\Popt}}{\waist_0 \waist_q}, 
\end{eqnarray} 
where $\waist_q$ denotes the width of the optical intensity distribution along the three directions ($q = x, y, z$) and $\waist_0^2 = \waist_x\waist_y$, $\Popt$ is  the optical power, $\rho$ the density of the particle.
Note that for tightly focused laser beams with linear polarization, as commonly used in optical trapping, the field distribution is slightly elongated along the direction of polarization of the incident field, which leads to slightly different trapping frequencies along the two transverse directions.
For larger oscillation amplitudes, the motion becomes anharmonic, and the nonlinear coefficients can be obtained from higher derivatives of the optical potential \cite{Gieseler2013}.

\subsubsection{Rotation}
For anisotropic particles, the light matter interaction is more complicated, since it depends upon the alignment of the object relative to the polarization axis of the field \cite{Stickler2016a, Kuhn2017}. 
For linearly polarized light, the particle experiences an optical torque which aligns the particle with respect to the polarization axis \cite{Geiselmann:2013gb}. For small deflections from the polarization axis the angular motion is harmonic.
In contrast to linearly polarized light, the polarization axis of circularly polarized light rotates at the optical frequency. This is too fast for the particle to follow. Nonetheless, light scattering transfers the angular momentum of the light to the particle and exerts a torque \cite{Kuhn2017}. The polarization anisotropy can originate from the intrinsic birefringence of the particle \cite{Arita:1ge, Bishop:2004ia, Paterson:2001ks} or from the anisotropic shape of the particle, e.g., a~cylinder \cite{Kuhn:2015fw} (c.f. Eqn.~\eqref{eqn:polarizability}).
The rotational degree-of-freedom of a levitated nanoparticle could be used to design microscopic engines in the classical and quantum regime \cite{Roulet2017}.
As of yet, there has been no such experimental implementation.

\subsection{Stochastic Forces}
So far we have only considered the static forces that are responsible for creating a potential landscape. In addition, stochastic forces excite the particle motion and lead to stochastic dynamics, which are of particular interest when studying thermodynamics of individual small particles.
The~stochastic forces result from the interaction of the particle with its environment. These~interactions lead to dissipation $\g{}^\noise$ and are the source of the different random forces acting on the particle (labelled $\noise$). The~strength of the random forces is characterized by their power spectral densities $\Sff{\noise}$. For most practical purposes, they can be considered as frequency independent (white noise). After a time $\approx 1/\g{CM}$, where $\g{CM} = \sum_\noise\g{}^\noise$ is the total damping rate, the particle reaches an~effective thermal equilibrium, which is characterized by the effective temperature through the fluctuation-dissipation~relation:
\begin{equation}
\label{eqn:temp_def}  
\T{CM} = \frac{\pi\Sff{} }{\kB m \g{CM} },
\end{equation}
where $\Sff{} = \sum_\noise\Sff{\noise}$ is the total force spectral density, $m=V \rho $ is the particle mass, and $\kB$ is Boltzmann's constant.
In the following we describe the individual contributions. They are:
 collisions with air molecules ($\noise =$ gas), radiation damping ($\noise =$ rad), feedback or cavity damping ($\noise =$ fb) and external driving ($\noise =$ drive).

\subsubsection{Gas Damping}
For high pressures, the interaction with the gas is so strong that the particle motion is heavily damped and its internal temperature $\T{int}$ and centre-of-mass temperature $\T{CM}$ quickly thermalize with the gas temperature $\T{gas}$. In this regime the damping becomes independent of pressure $\g{CM}/2\pi\approx 3 a \viscosity/m$, as predicted by Stokes law, where $\viscosity$ is the viscosity. However, for strong absorbers, e.g., resonantly illuminated plasmonic particles \cite{Baffou:2012iz, Baffou:2010id, Jauffred2015}, the particle's internal temperature can rise significantly above the temperature of the environment, leading to so called ``hot Brownian motion'' \cite{Rings2010, Rings2012, Falasco2014}.

For decreasing pressure, the mean free path of the gas molecules increases (e.g., $\bar{l}\sim 60\,\upmu$m at 1\,mbar). As a consequence, the particle no longer thermalizes with the gas since the impinging gas molecules no longer carry away enough thermal power to balance the optical absorption from the trapping laser. Due to the increased internal temperature $\T{int}$ of the particle, the average energy of the gas molecules after a collision with the particle increases. The process by which a surface exchanges thermal energy with a gas is called accommodation, which is characterized by the accommodation coefficient
$\acccoeff = (\T{em} - \T{gas})/(\T{int} - \T{gas})$,
where $\T{em}$ is the temperature of the gas molecules emitted from the particle surface. Accommodation quantifies the fraction of the thermal energy that the colliding gas molecule removes from the surface, such that $\acccoeff =1$ means that the molecule fully thermalizes with the surface. Since the mean free path in a dilute gas is long, one can safely assume that an emitted molecule will not interact again with the particle before thermalizing with the environment. Consequently, we can consider the particles that impinge on the particle surface and those that leave the surface as being in equilibrium with two different baths with temperatures corresponding to the temperature of the environment and the particle surface, respectively \cite{Millen2014}. Therefore, we get an~additional contribution to the damping from the emerging hot molecules
\begin{equation}\label{eq:gas_damping_2bath}
  \frac{\g{em}}{2\pi} = \frac{1}{16}\sqrt{\frac{\T{em}}{\T{gas}}}\g{gas} ,\quad
  \Sff{em} =
 \frac{m\kB}{\pi}\left[\acccoeff\T{int} +(1-\acccoeff\right)\T{gas}]\g{em}.
\end{equation}

For pressures below $\Pg' =0.57 \kB \T{gas} \left/\crosssectiongas a\right.\approx 54.4\,{\rm mbar} \times (a/ \upmu {\rm m})^{-1}$, where the mean free path $\bar{l} = \kB \T{gas}\left/(\sqrt{2}\crosssectiongas\Pg)\right.$ is much larger than the radius of the particle $a$, the damping is linear in the pressure $\Pg$ and given by 
\begin{equation}\label{eq:gas_damping_lin}
 \frac{\g{gas}}{2\pi} =  \frac{3}{\pi\sqrt{2}}\frac{\viscosity\crosssectiongas}{\kB \T{gas} \rho}\frac{\Pg}{a}, 
\end{equation}
where $\viscosity = 2\sqrt{\mgas \kB\T{gas}}\left/ 3\sqrt{\pi} \crosssectiongas \right.$ is the viscosity of a dilute gas, $\crosssectiongas$ is the cross-section of the air molecules and $\mgas$ the molecule mass. The total damping due to the hot particle with the gas environment is $\g{em}+\g{gas} =2\pi \cg \Pg / a$, where typically $\cg\approx   50\, \Hz (\mum/ \mbar)$.

So far, we have only considered spherical particles. For anisotropic particles, e.g., a rod, the~friction term is different along each of the axes, and depends upon the alignment of the particle. As~a~consequence, the friction coefficient has to be replaced by a tensor $\boldsymbol{\Gamma}$ and the damping in a direction $\mathbf{s}$ is given by $\boldsymbol{\Gamma}\cdot \mathbf{s}$. For a thorough discussion see Ref.~\cite{Martinetz2018}.

\subsubsection{Radiation Damping}
\label{sec:photon}

At very low pressure ($\leq 10^{-6}\,\rm mbar$), gas damping becomes extremely small and photon shot noise starts to dominate \cite{Jain2016a}. Photon shot noise is a consequence of the particulate nature of light. \mbox{As a consequence}, photons arrive at discrete times, where the number of photons arriving per time interval $\Delta t$ is given by $\sqrt{\Delta t \Popt\left/\hbar \wopt \right.} $. The recoil from the fluctuating number of photons impinging on the nanoparticle can be modeled as an effective bath with the characteristics \cite{Novotny:2017bd}
\begin{equation}\label{eq:radiation_damping}
  \frac{\g{rad}}{2\pi} = \dpcoeff\frac{\Pscat}{2\pi m c^2}
\quad{\rm and}\quad
  \Sff{rad} = \dpcoeff \frac{\hbar \omega \Pscat}{2\pi c^2},   
\end{equation}
where $\dpcoeff$ depends on the direction of motion of the particle with respect to the polarization of the laser and is $\dpcoeff=2/5$ for motion along the direction of polarization and $\dpcoeff=4/5$ for motion perpendicular to the polarization. The scattered power is $\Pscat = \scatcross \Iopt$, where $\scatcross = |\alpha|^2\kopt^4/6\pi\epsilon_0^2$ and $\Iopt$ is the laser intensity. The effective temperature of this bath can be calculated via Eqn.~\eqref{eqn:temp_def}.

\subsubsection{Artificial Damping and Heating}
The noise processes described so far are present in any experiment with optically levitated nanoparticles in high vacuum. In addition, random forces and damping can be introduced through external fields that are under experimental control. Importantly, since energy can be injected or extracted from the particle, i.e., it is not in thermal equilibrium, the fluctuation-dissipation relation does not have to hold and the effective damping and temperatures can be controlled independently.
Feedback cooling damps the particle motion at a rate $\g{fb}$ without adding any fluctuating forces, thus $\Sff{fb}=0$ and it is therefore referred to as cold damping \cite{Gieseler:2012bi, Li:2011jl}. Note that this simplified picture assumes that the feedback signal is perfect and that it does not feedback any noise, which in general is not true. Similarly, cavity cooling up-converts the particle energy to optical frequencies, which are effectively at zero temperature because $\hbar \omega \gg \kB \T{env}$ in a room temperature environment \cite{Kiesel:2013bp, Chang2010, RomeroIsart2010}. Conversely, fluctuations of the trapping or control fields only add fluctuating forces without providing damping. Hence, $\g{drive} = 0$ and $\Sff{drive} = \q^2 S_{\rm qq}$, where $\q$ is the coupling parameter to the control field and $S_{\rm qq}$ its spectral density.
This can be realized for example with fluctuating electric fields, where $\q$ corresponds to the charge on the particle \cite{Frimmer:2017jk, Mestres:2014bx, Martinez:2013fe}.

Generally, one has to be careful to define a temperature for a system out of equilibrium \cite{CasasVazquez2003}. However, the situation we present here is somewhat simple due to it being steady-state and for many practical purposes the effective bath model that is characterized by an effective damping/temperature gives a good description. However, one can also create situations where this is no longer true. For instance, by parametric feedback damping, the temperature alone is not sufficient to give a full description of the bath \cite{Gieseler2014}. 

\section{Brownian Motion}
\label{sec:Brownian}
As discussed in the introduction, 
the Brownian particle serves as an exemplary model to describe a variety of stochastic processes in many fields, including physics, finance and biology. 
Therefore, a~particle trapped in optical tweezers is paradigmatic since it is a direct experimental realization of the idealized Brownian particle. 
Brownian motion in nonequilibrium systems is of particular interest because it is directly related to the transport of molecules and cells in biological systems \cite{Gnesotto:2017ul}. Important examples include Brownian motors, active Brownian motion of self-propelled particles, hot Brownian motion, and Brownian motion in shear flows \cite{Bechinger:2016cf}. Recent theoretical studies also found that the inertia of particles and surrounding fluids can significantly affect the Brownian motion in nonequilibrium systems \cite{Nicolis2017}.

In this section we will discuss the basics of Brownian motion. We will mainly treat the aspects that are necessary for understanding the following discussion of thermodynamics with levitated nanoparticles. For details on the theory of Brownian motion we refer the reader to the work of Ornstein, Uhlenbeck and Wang \cite{Uhlenbeck:1930tn, Wang:1945wd} and for a recent review on Brownian motion in the underdamped regime we refer the reader to Ref.~\cite{Li2013}.

\subsection{Harmonic Brownian Motion}
Under the influence of trapping forces, the particle will be localized about its equilibrium position. For small displacements, the trap can be approximated by a three-dimensional harmonic potential.
The three motional degrees of freedom are largely decoupled and we limit the discussion to a single coordinate $q(t)$ ($q = x,y, z$).
The equation of motion for a harmonically trapped Brownian particle is~\cite{Wang:1945wd}
\begin{equation}\label{eq:HarmonicLangevin}
\ddot{q} \,+\, \g{} \!\; \dot{q} +\wo^2 q=\;   \sqrt{2\kB \T{CM}\, \g{CM}/m}\,\whn(t). 
\end{equation}

The particle oscillates in the trap at the characteristic frequency $\tilde{\w} = \sqrt{\wo^2-\g{CM}^2/4}$. For the optical potential the trap frequency $\wo$ is given by Eqn.~\eqref{eqn:frequencies_translational}.
We distinguish between three cases, the overdamped ($\wo \ll \g{CM}$), the critically damped ($\wo \approx \g{CM}$) and underdamped case $\wo \gg \g{CM}$. This stochastic equation of motion has been studied in detail by Ornstein and Uhlenbeck \cite{Uhlenbeck:1930tn} and we summarize their results here. The variance of the position of a Brownian particle in an under-damped harmonic trap is
\begin{equation}\label{eq:pos_variance_underdamped}
\var{q}(t)=  \frac{2\kB \T{CM}}{m\wo^2}\left[1-e^{-\frac{1}{2}\g{CM} t}\left(\cos(\tilde{\w} t) +\frac{\g{CM}}{2\tilde{\w}}\sin(\tilde{\w} t)\right)\right].
\end{equation}

In the over-damped harmonic trap, set $\tilde{\w}\to i \tilde{\w}$. In a critically damped harmonic trap, set $\tilde{\w}\to 0$. The position autocorrelation function is related to the variance as {follows}

\begin{subequations}
\begin{equation}\label{eq:autocorrelation}
\corr{q}{q}
   = \frac{\kB \T{CM}}{m\wo^2} - \frac{1}{2}\var{q}(t),
\end{equation}
and the velocity autocorrelation and the position-velocity correlation function are{ given by }

\begin{align}\label{eq:vel_corr}
\corr{v}{v}&=  \frac{\kB \T{CM}}{m}e^{-\frac{1}{2}\g{CM} t}\left(\cos(\tilde{\w} t) -\frac{\g{CM}}{2\tilde{\w}}\sin(\tilde{\w} t)\right),\\ \label{eq:velpos_corr}
\corr{q}{v}&=\corr{v}{q}=  \frac{\kB \T{CM}}{m \tilde{\w}}e^{-\frac{1}{2}\g{CM} t}\sin(\tilde{\w} t).
\end{align}
\end{subequations}

\begin{figure}[h]
\begin{center}
  \includegraphics[width=15 cm]{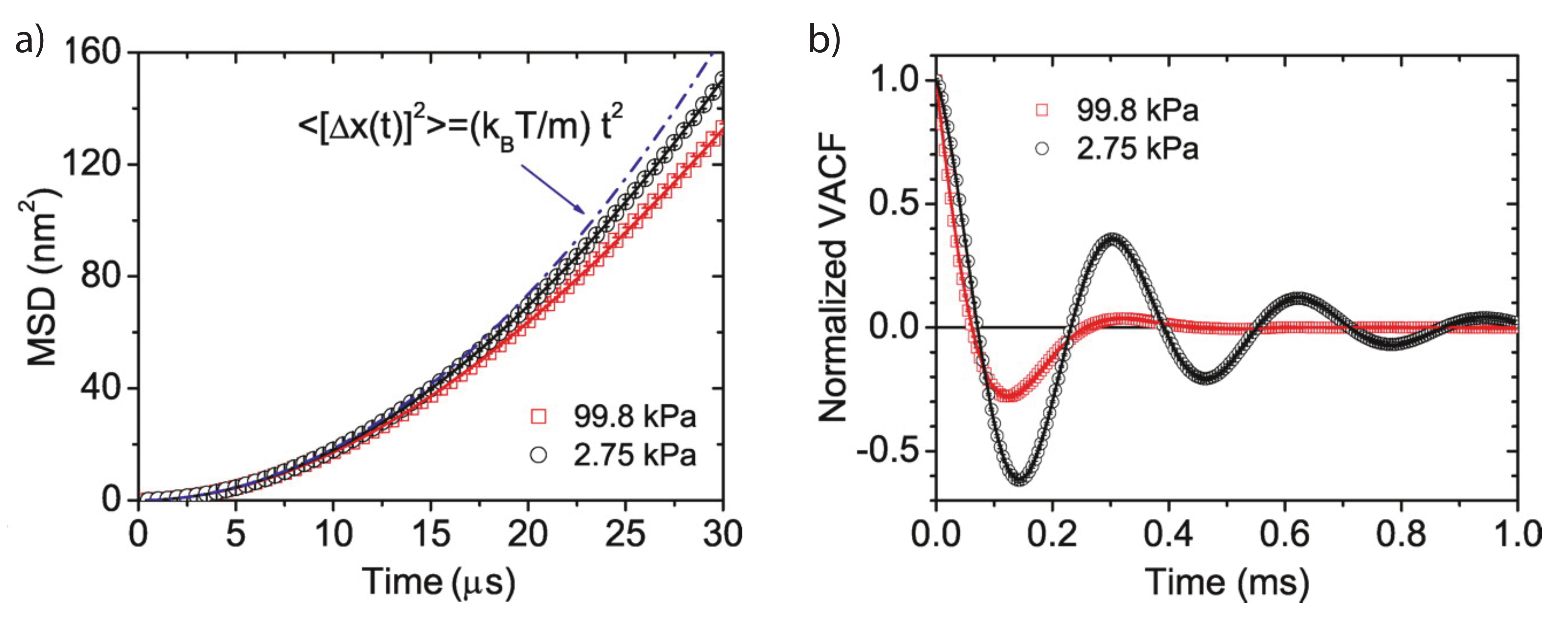}
  \caption{{{First experimental observation of the instantaneous velocity of a Brownian particle.}}
  (\textbf{a})~The mean-square displacement for short times is proportional to $t^2$, a signature of ballistic motion.
  (\textbf{b})~The normalized velocity autocorrelation functions for different pressures in perfect agreement with Eqn.~\eqref{eq:vel_corr}. 
  Figures taken from \cite{Li2010} with permission from Science.
   \label{fig:BrownianMotion}}\vspace{-6pt}
\end{center}
\end{figure}

For a long time it was believed that the timescale at which these correlations exist is too fast to be observable in experiment \cite{Einstein:1905gw}. 
The first experimental observation was first achieved in vacuum~\cite{Li2010} and later in liquid \cite{Kheifets:2014hq}, demonstrating that levitated nanoparticles indeed allow one to attain an~entirely new parameter regime to study thermodynamics of individual particles.
Fig.~\ref{fig:BrownianMotion} shows the experimental results from Li et al. \cite{Li2010}.

\subsection{Power Spectral Density and Calibration}
\begin{figure}[h]
\begin{center}
	\includegraphics[width=0.9\textwidth]{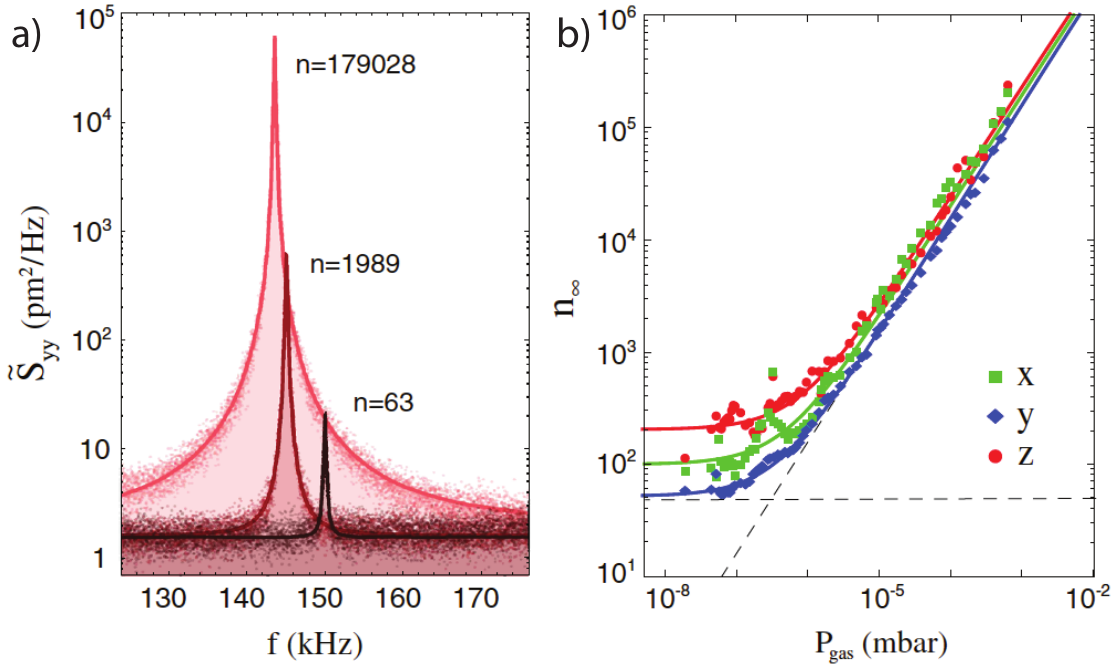}
  \caption{{{Feedback cooling of a levitated nanoparticle}}.
  (\textbf{a}) Power spectral density under phase locked feedback cooling at three different pressures and constant feedback gain. The area under the power spectral densities is a measure for the effective center-of-mass temperature.
  (\textbf{b}) The effective temperature expressed in terms of the phonon occupation as a function of gas pressure.
  Figures~reproduced from Physical Review Letters \cite{Jain2016a}.
   \label{fig:Cooling}}
\end{center}
\end{figure}

According to the Wiener-Khinchin theorem, the position autocorrelation function is the Fourier transform of the power spectral density $S_{qq}(\w) = \int_{-\infty}^\infty \corr{q}{q}e^{i\w t}\d t$, which for Eqn.~\eqref{eq:HarmonicLangevin} is given~by 
\begin{equation}\label{eq:PSD_position}
  S_{qq}(\w) = |\chi(\w)|^2 \Sff{}(\w)
  = \frac{\g{CM} \kB \T{CM}\left/\pi m\right.}{(\w^2-\wo^2)^2+\g{CM}^2\w^2},
\end{equation}
where 
$
  \chi(\w) = m^{-1}\left[\w^2-\wo^2+i\g{CM}\w\right]^{-1}
$
is the response function or susceptibility of a harmonic oscillator. In the underdamped regime, the frequency spectrum of the autocorrelation function is strongly peaked around the trap frequency $\wo$, whereas when overdamped the frequency spectrum is broad.
For an example of the power spectral density in the underdamped regime see Fig.~\ref{fig:Cooling}a.

The power spectrum in Eqn.~\eqref{eq:PSD_position} is valid for a harmonic oscillator. However, the actual trapping potential is nonlinear.
For a symmetric potential the lowest order nonlinear term is a cubic or Duffing nonlinearity. In an optical trap the symmetry is broken along the direction of propagation of the trapping laser and along the vertical direction due to the scattering force and gravity, respectively. However, the symmetry breaking tends to be small and therefore the quadratic nonlinearity is typically neglected.
Because of the Duffing nonlinearity, the oscillation frequency along an axis $i = x,y,z$ becomes a function of the oscillation amplitude and is red shifted by \cite{Gieseler2013}
$
  \Delta\w_{\rm i} =\frac{3}{8}\w_i \sum_j \xi_{ij}  \a_{\rm j}^2
$
where $\a_j$ is the instantaneous amplitude of mode $j$ and the $\xi_{ij}\sim 1/\waist_j^2$ are the Duffing terms, where $\waist_j$ is the width of the optical potential along the $j$ direction.
The~frequency shift due to changes in the oscillation amplitudes is also known as self-phase modulation ($j=i$) and cross-phase modulation ($j\neq i$).
The~frequency shift can be neglected as long as it is much smaller than the linewidth $\g{CM}$. This is the case for high pressure $\simeq$1 $\mbar$, but for low pressures the amplitude fluctuations of the particle lead to significant frequency fluctuations and the power spectral density becomes distorted.
This nonlinear Brownian motion typically does not play a role in nanomechanical systems because the amplitude fluctuations are small. 
However, levitated nanoparticles have a low mass and a~high motional quality-factor and, therefore, nonlinear Brownian motion can be observed in these systems~\cite{Gieseler2013}.

The power spectral density is a useful tool in experiments with harmonic oscillators, since the dynamics of the oscillator can be separated from (spectrally distant) noise. In addition, the analysis of the power spectral density allows one to extract the center-of-mass temperature of the oscillator and the damping rate \cite{Li2010, Gieseler:2012bi, Millen2014, Hebestreit2018}.
However, to avoid miss-calibration due to the above mentioned nonlinearities, one should use the velocity power spectral density $S_{\dot{q}\dot{q}} = \w^2 S_{qq}$ instead of \eqref{eq:PSD_position}
\cite{Hebestreit:2017wh}.

\subsection{Quantum Brownian Motion}
In the quantum regime, when the center-of-mass temperature is of the order of a single quantum of motion $\kB \T{CM}\approx \hbar \w_0$, the position autocorrelation Eqn.~\eqref{eq:autocorrelation} contains the product of the Heisenberg time-evolved operators $\hat{q}(t)$, $\hat{q}(0)$, which do not commute.
As a result, the spectrum \cite{Clerk:2010dh}

\begin{equation}
	\label{eq:PSD_quantum}
  S_{Q}(\w) %
  = \frac{\hbar/\pi}{1-\exp\left(-\frac{\hbar \w}{\kB \T{CM}}\right)} {\rm Im} \chi(\w)
= \frac{\hbar\w m\g{CM}/\pi}{1-\exp\left(-\frac{\hbar \w}{\kB \T{CM}}\right)} |\chi(\w)|^2,
\end{equation}
is asymmetric in frequency and the PSD at positive frequencies is a factor $\exp(\hbar \wo\left/\kB \T{CM}\right.)$ higher than the PSD at negative frequencies. The positive-frequency part of the spectral density is a measure of the ability of the oscillator to absorb energy, while the negative-frequency part is a measure of the ability of the oscillator to emit energy.
Therefore, we can understand the positive frequency part of the  spectral density as being related to stimulated emission of energy into the oscillator, while the negative-frequency part is related to the emission of energy by the oscillator.

Typically, the motional frequencies of a levitated particle are $\sim$100 kHz. Hence, the ground-state temperature is a few micro-kelvin and therefore out of reach for cryogenic techniques, and one has to resort to active cooling methods. Recent experiments using feedback cooling have already attained motional occupations of a few tens of phonons \cite{Jain2016a}. However, a measurement of the sideband asymmetry in a homodyne measurement, as observed in other nano-mechanical systems \cite{SafaviNaeini:2012ih, Weinstein:2014fc, Underwood:2015hd, Peterson:2016ka, Kampel:2017gt}, is~still elusive.

\section{Trap Stability and Kramers Turnover\label{sec:Kramers}}

For a particle to be trapped in optical tweezers, the axial component of the gradient force must exceed the destabilizing effects of the scattering force and gravity. The scattering force is negligible for small particles but increases quickly with particle size such that large particles are pushed away from the focal volume. In addition, the ratio between the scattering and gradient forces scales with the refractive index contrast \cite{Spesyvtseva2016}. This places an upper limit on the maximum particle size and materials that can be trapped, even if they experience very little optical absorption. The destabilizing effect from the scattering force can be circumvented by using a configuration with counter-propagating beams~\cite{Divitt:2015gw, Li2010}. However, in this case polarization fluctuations translate into intensity fluctuations, which can also destabilize particles in the trap. Besides, radiometric forces can play a role in the stability condition for larger particles ($\sim$$\upmu$m), where non-uniform heating leads to temperature gradients across the particle \cite{Millen2014} and the resulting forces might destabilize the trap \cite{Ranjit:2015bb}.

Overcoming the destabilizing effects from the scattering force is not sufficient to guarantee that particles can be trapped. As we discussed earlier, the particle is subject to fluctuating forces from the environment.
The energy of a thermal bath follows a Maxwell-Boltzmann distribution with mean value of $\kB \T{CM}$. Since the tail of the distribution extends to high energies, the potential depth should be at least $\approx 10 \kB \T{CM}$ \cite{Ashkin1986} to make particle escape through thermal excitation unlikely (the likelihood of finding the particle with energy $\approx 10 \kB \T{CM}$ is less than $0.02\%$).
Hence, there is a finite probability that the particle will gain enough energy to escape the potential, even when it is confined by a potential much deeper than $\kB\T{CM}$, in a process known as Kramers escape. This form of ``\emph{classical tunneling}'' appears in a diverse range of physical systems, including chemical reaction rates, protein folding, atomic transport in optical lattices and molecular diffusion at solid-liquid interfaces.
The Kramers escape rate is given by an Arrhenius law
\begin{equation}\label{eq:KramersRate}
  \R{K} = \R{0} \exp\left(-\frac{U_{\rm barrier}}{\kB \T{CM}}\right)
\end{equation}
where $\R{0}$ is the attempt frequency and $U_{\rm barrier}$ is the barrier height. From the Boltzmann factor in Eqn.~\eqref{eq:KramersRate} it follows that such a transition is exponentially suppressed if the potential is much deeper that the thermal energy $U_{\rm opt}\gg \kB \T{CM}$.
Kramers found \cite{Kramers1940} that in the underdamped regime, the transition rate increases with \emph{increasing} friction, and that in the overdamped regime the transition rate increases with \emph{decreasing} friction, with the transition region labelled the turnover. Fifty years later, a theory was developed that linked the two regimes \cite{Melnikov1991, Hanggi:1990ud}.

Closely related to Kramers escape is the Kramers turnover problem. It describes the transitioning between two local potential minima as the friction is varied. This is often more relevant in physical situations, describing the transitions between two protein configurations, for example.
It is also much more convenient to study experimentally, since the particle is not lost after the transition but instead recaptured in the other well.
In particular, levitated particles are well suited to studying the Kramers escape and recently led to its first quantitative observation \cite{Rondin2017}.
The double well potential can be created by using two tightly focused laser beams.
The intensity and exact relative position of the two foci determines the height of the barrier. The hopping rates between the two wells is determined by the local curvatures of the potential at the extrema and by the interaction strength with the environment.
In addition, the interaction strength can be varied over many orders of magnitude through a change in the gas pressure $\Pg$.

\begin{figure}[h]
\begin{center}
	\includegraphics[width=0.7\textwidth]{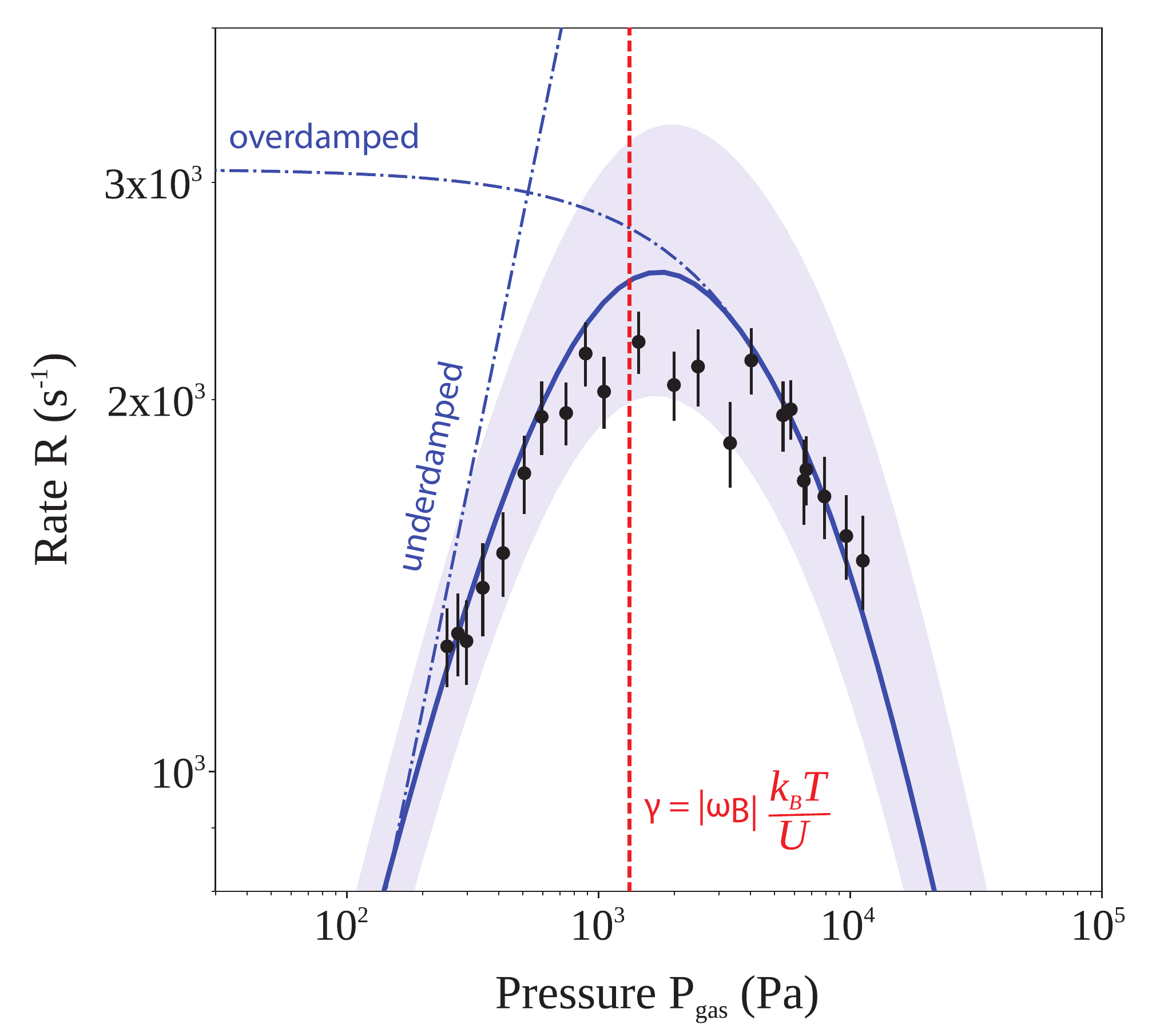}
  \caption{{{Measurement of the Kramers turnover with a levitated nanoparticle.}}
Data illustrating the first experimental observation of Kramers turnover, taken from \cite{Rondin2017}. The full theory from \cite{Melnikov1991} (solid line) is shown as a solid blue line together with the limiting cases as predicted by Kramers \cite{Kramers1940} (dot-dashed lines). 
The red dashed line highlights the expected turnover point as predicted from the measured shape of the double well potential and is in excellent agreement with the experimental observations.
   \label{fig:Kramers}}
\end{center}
\end{figure}

Fig.~\ref{fig:Kramers} shows the experimental data from Rondin {et al.}, which for the first time measured the Kramers rate across the turnover \cite{Rondin2017}, using an optically levitated nanoparticle.
In addition, the~figure includes the limiting cases in the high and low damping regime, and the full solution for arbitrary~damping. 

\section{Effective Potentials in the Steady State\label{sec:EffectivePotential}}

At constant trapping laser power, the trapping potential is static in time $U_{\rm opt}(\mathbf{r})$. However,~through modulation of the trapping beam intensity, the optical potential becomes time-dependent. This is particularly useful when studying non-equilibrium dynamics and for engineering effective baths in the context of nano heat engines.
From Eqn.~\eqref{eqn:frequencies_translational} it follows that a change in optical power $\delta \Popt(t)$ changes the trap frequency by $\w(t) = \wo(1+\emod(t)/2)$, where $\emod(t) = \delta \Popt(t) \left/ \Poptm\right.$ and $\Poptm$ is the mean optical power.
Energy is most effectively exchanged between the trapping laser and the particle if the modulation $\emod(t) = \emod_0\cos(\omegamod t)$ occurs at twice the trapping frequency $\omegamod\approx 2\w_0$. The~flow of energy is thereby determined by the relative phase $\phimod$ between the particle oscillation and the laser intensity modulation (note that $\phimod$ does not appear in $\emod(t)$, since the modulation serves as the time reference and $\phimod$ is the phase of the particle with respect to the modulation). If the modulation is in-phase, energy is extracted (cooling), while the motion is excited when the modulation is out-of-phase (heating).
Without active stabilization of the modulation phase with respect to the particle motion, the relative phase is random. However, since the out-of-phase motion is amplified and the in-phase motion is damped, the out-of-phase motion quickly dominates and the particle motion synchronizes or entrains with the parametric modulation \cite{Gieseler:2014wt}.

Therefore, to achieve cooling the phase needs to be actively stabilized, for instance with a~phase-locked loop \cite{Jain2016a}.
Alternatively, a parametric feedback modulation of the form $\emod_\text{fb}(t) = -(\fb/\wo) q(t)\dot{q}(t)$ \cite{Gieseler:2012bi}, where $\fb$ parameterizes the feedback strength, leads to a modulation at the parametric resonance condition, ensuring a phase that is optimized for extracting energy from the mechanical mode.
In contrast to feedback with a phase-locked loop, where the modulation amplitude is constant, here the modulation amplitude is proportional to the particle energy. As a consequence, the particle feels a nonlinear friction force with $\g{NL}\propto E$.
This kind of friction has also be observed in carbon based nanomechanical resonators \cite{Eichler:2011fi} and leads to non-equilibrium steady states that can no longer be described by a thermal distribution as we will discuss in the following.

\subsection{Effective Potential for the Energy}
The main physics of the particle motion under parametric modulation and feedback is captured by a one dimensional equation of motion along each axis: 
\begin{equation}\label{eq:ParametricLangevin}
\ddot{q} \,+\, \g{} \!\; \dot{q} +\wo^2\left[1+\emod_0 \cos(\omegamod t)+\xi q^2 +\wo^{-1}\eta q\dot{q} \right]q=\;   \sqrt{2\kB \T{CM}\g{CM}/m}\,\xi(t),
\end{equation}
where  $q = (x, y, z)$.
The total energy of a single degree-of-freedom is given by
\begin{equation}\label{eq:Energy}
E(q,p)=\frac{1}{2}m\wo^2 q^2+ \frac{p^2}{2m}+\frac{1}{4}\xi m\wo^2 q^4,
\end{equation}
where $p(t) = m\dot{q}$ is the momentum.
The energy obeys the stochastic differential equation \cite{Gieseler:2015bl}
\begin{equation}
{\rm d} E = \left[-\g{CM} (E - \kB \T{CM})
 -\frac{\eta  \wo E^2}{2m \w^2}
-\frac{E\emod\wo^2 \sin(2\phimod)}{2\w}
\right]{\rm d}t 
+\sqrt{2E \g{CM} \kB \T{CM}}{\rm d}W. 
\label{equ:SDE_energy}
\end{equation}

From \eqref{equ:SDE_energy} one derives the probability distribution for the energy
\begin{equation}
P_E(E) = \frac{1}{Z} \exp\left\{-\beta H(E)\right\},
\label{eq:energy_distribution}
\end{equation}
where $Z=\int P_E(E){\rm d}E$.
Thus, the energy distribution is that of an equilibrium system with effective~energy
\begin{equation}\label{eq:Heff}
\Heff=\left[1+\frac{\emod_0\wo^2 \sin(2\phimod)}{2\g{CM}\w}\right]E+\frac{\eta  \wo}{4m\g{CM} \w^2}E^2.
\end{equation}

While the term proportional to $E^2$ is caused by the feedback cooling, the term proportional to $E$ is affected only by the parametric modulation. 
Note that the Duffing term is included in the energy $E$ on the right-hand side of the above equation (c.f. Eqn.~\eqref{eq:Energy}).

Eqn.~\eqref{equ:SDE_energy} is quite general, as it captures the dynamics of the slowly varying energy under parametric heating without active stabilization, phase-locked loop feedback cooling and parametric feedback cooling.
However, one has to bear in mind that the oscillation frequency $\w$ is not necessarily the same as the frequency $\wo$ of the unperturbed harmonic oscillator. For instance, for strong modulation the particle motion entrains with the modulation and $\w \approx \omegamod/2$ \cite{Gieseler:2014wt}, while for weak modulation $\w \approx \wo$. 
The weak and strong regime are determined by the threshold condition
$
\emod_0 > 2\Qf^{-1}\sqrt{1+\Qf^2\left(2-\omegamod/\wo \right)^2}\approx 2\Qf^{-1}
$, the approximation being exact at parametric resonance $\omegamod = 2\wo$ and $Q = \wo/\g{CM}$ is the quality factor.
Above threshold, the effective temperature diverges and the motion transitions from a thermal state to a coherent oscillation, similar to the lasing condition of an optical oscillator \cite{Gieseler:2014wt}.

Note that the evolution of the position of a {\emph{real}} Brownian particle in the underdamped regime and in a {\emph{time-dependent}} optical potential is determined by Eqn.~\eqref{eq:ParametricLangevin}.
In contrast, Eqn.~\eqref{equ:SDE_energy} describes the evolution of its energy or amplitude. 
This evolution can be interpreted as the evolution of an~underdamped, albeit {\emph{fictitious}}, Brownian particle in a {\emph{static}} potential \cite{Gieseler:2015bl}.
Interestingly, this {\emph{fictitious}} Brownian particle can exhibit dynamics similar to what we have seen for the {\emph{real} }Brownian particle, such as transitions between two local minima that are described by Kramers' theory \cite{Ricci:2017eh}.

\subsection{Effective Temperature}
Without non-linear parametric feedback ($\eta = 0$), the energy distribution is that of a harmonic oscillator with effective temperature 
\begin{equation}\label{eq:Teff}
\T{CM}' = \T{CM}\left(1+\frac{\emod_0\wo^2 \sin(2\phimod)}{2\g{CM}\w}\right)^{-1}.
\end{equation}

Eqn.~\eqref{eq:Teff} states that parametric modulation of the trapping potential results in an effective temperature change of the environment, where the particle centre-of-mass temperature changes from $\T{CM}$ to $\T{CM}'$. For $-\pi/2<\phimod<0$, $\T{CM}'>\T{CM}$, that is the particle motion is heated, while~for $0<\phimod<\pi/2$, $\T{CM}'<\T{CM}$ and the particle motion is cooled. The rate at which the particle thermalizes with this effective bath is $\g{CM}'   = \g{CM} \left(\T{CM}/\T{CM}'-1\right)$, where the largest rates are achieved at $\phimod = -\pi/4$ and $\phimod = \pi/4$, for heating and cooling respectively. If the relative phase between the particle motion and the modulation $\phimod$ is not stabilized actively, the particle motion will self-lock to $\phimod = -\pi/4$. Thus, an effective hot bath can be implemented easily by a simple modulation of the trapping laser at $\omegamod\approx 2\wo$.
Fig.~\ref{fig:Cooling} shows the effective temperature or occupation number for a~particle under high vacuum as a function of pressure (i.e., $\g{CM}$) and constant feedback strength ($\emod_0$).

\section{Relaxation \label{sec:Relaxation}}
In the steady-state, a trapped particle samples the distribution Eqn.~\eqref{eq:energy_distribution}, which depends on experimental parameters, such as the average power of the trapping laser, and the strength and frequency of the modulation of any potential modulation. Hence, under a non-adiabatic change of the parameters, the systems relaxes into a new steady state.
The Fokker-Planck equation that describes the time for the evolution probability density function $P_E(E, t)$, including feedback and modulation, is~given by \cite{Gieseler:2015bl}
\begin{small} 
\begin{equation}\label{eq:FokkerPlanckEnergy}
\frac{\partial P_E(E, t)}{\partial t}=\frac{\partial }{\partial E}\left[\g{CM} (E - \kB \T{CM})
 +\frac{\eta  \wo E^2}{2m \Omega^2}
 +\frac{E\emod_0\wo^2 \sin(2\phimod)}{2\w}\right]P_E(E, t)
 +{\g{CM} \kB \T{CM}}\frac{\partial^2 }{\partial E^2}E P_E(E, t).
\end{equation}
\end{small}


In general it is non-trivial to find an analytic solution to Eqn.~\eqref{eq:FokkerPlanckEnergy}.
Amazingly, in the absence of feedback cooling ($\fb = 0$), the equation of motion for the energy corresponds to the 
Cox-Ingersoll-Ross model for interest rates, for which the exact analytical solution is given by the Noncentral Chi-squared distribution \cite{Salazar:2016ey}
\begin{equation}
  P_E(E|E_0, t) = c_t e^{-c_t(E+E_0e^{-\g{CM} t})}I_0\left(2c_t\sqrt{E E_0 e^{-\g{CM} t}}\right),
  \end{equation}
where $c_t = \beta\left(1-e^{-\g{CM} t}\right)^{-1}$, $I_0(x)$ is the modified Bessel function of the first kind and $E_0$ is the initial energy, i.e., $P_0(E|E_0) = \delta(E-E_0)$. As expected, the equilibrium distribution   $P_\infty(E|E_0) = \beta\exp(-\beta E)$ does not depend on the initial conditions and is given by the Maxwell-Boltzmann distribution at temperature $\T{CM} = 1/(\kB \beta)$.
If the system is initially prepared at $t = 0$ in a steady state with energy distribution $P_0(E_0)$, the energy distribution after time $t$ is 
\begin{equation}\label{eq:Relaxation}
  P_E(E, t) = \int_0^\infty P_E(E|E_0, t)P_0(E_0)\d E_0.
\end{equation}

For an initial Maxwell-Boltzmann distribution, corresponding to a thermal equilibrium distribution at temperature $\T{0}$, the energy distribution at time $t$ is also a Maxwell-Boltzmann distribution

\begin{equation}
  P^\text{MB}_E(E, t) = \beta(t) e^{-\beta(t)E},
\end{equation}
with time dependent temperature 

\begin{equation}
  \T{CM}(t) = \T{\infty}+(\T{0}-\T{\infty})e^{-\g{CM} t}.
\end{equation}

\begin{figure}[h]
\begin{center}
  \includegraphics[width=\textwidth]{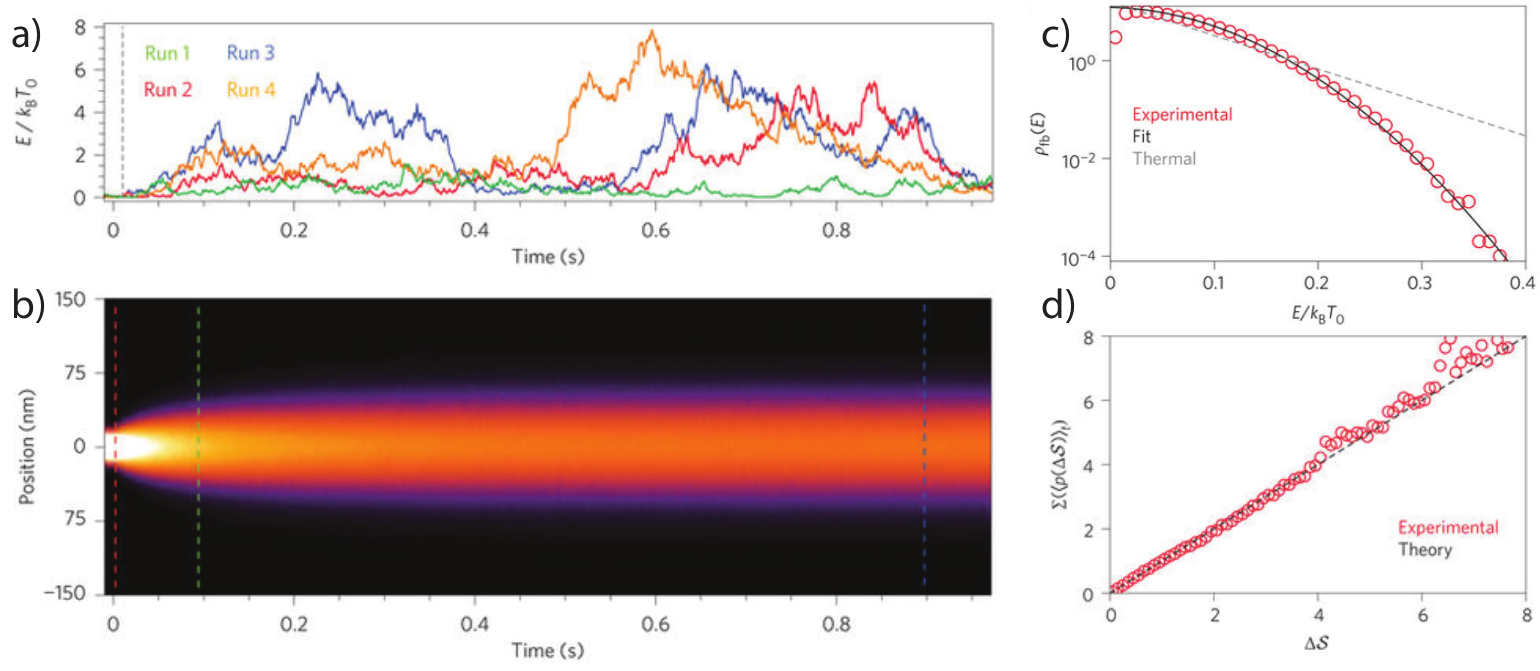}
  \caption{{{Relaxation from a non-equilibrium steady state}}.
  (\textbf{a}) Individual trajectories of the energy as the system relaxes toward equilibrium
  (\textbf{b}) Position distribution during the relaxation process. The~energy distribution is given by Eqn.~\eqref{eq:Relaxation}.
  (\textbf{c}) Energy distribution in the steady state ($t\leq 0$) in agreement with Eqn.~\eqref{eq:energy_distribution}. The deviation from a thermal state due to the nonlinear feedback is clearly visible.
 (\textbf{d}) Experimental verification of the detailed fluctuation theorem (c.f. Eqn.~\eqref{eq:DetailedFluctuationTheoremRelEntropy}).
  All~figures taken from \cite{Gieseler2014} with permission from  Nature Nanotechnology.
   \label{fig:FluctuationTheorem}}
\end{center}
\end{figure}

Note that the initial temperature $\T{0}$ and final temperature $\T{\infty}$ can be controlled in the experiment by modulation of the trapping laser, as discussed earlier. Explicitly, a levitated nanoparticle can be cooled via feedback to a centre-of-mass temperature $\T{CM}$ far below the ambient temperature. Once the feedback modulation is switched off, the particle will thermalize with the environment (in general via collisions with surrounding gas), at an average rate $\g{CM}$, which can be controlled by varying the gas pressure. Fig.~\ref{fig:FluctuationTheorem} shows the relaxation from a non-equilibrium state towards thermal equilibrium.
The rate at which the particle reaches equilibrium can be accelerated using time-dependent potentials. This has been demonstrated recently with a colloidal particle \cite{Martinez2016} and a similar strategy has been proposed for underdamped systems \cite{Chupeau:2018ww}.

\section{Fluctuation Theorems\label{sec:FluctuationTheorems}}

As a system relaxes to a thermal equilibrium the dynamics satisfy detailed balance with respect to the equilibrium distribution. The time reversibility of the underlying dynamics implies that the Crooks-like \cite{Crooks:377970,Collin:2005fx} transient fluctuation theorem \cite{SeifertReview, Gieseler2014}
\begin{equation}\label{eq:DetailedFluctuationTheoremRelEntropy}
  \frac{P(-\dS)}{P(\dS)} = e^{-\dS},
\end{equation}
for the relative entropy change $\dS = \beta \Q+\Delta \entropy$ (or Kullback-Leibler divergence) holds. The quantity $\Delta \entropy = \entropy(t)-\entropy(0)$ is the difference in trajectory dependent entropy $\entropy(t) = -\ln P_0(u(t)) $ between the initial and the final states of the trajectory and $\Q$ is the heat absorbed by the bath at reciprocal temperature $\beta$.
Here, $u(t)$ denotes an entire trajectory of length $t$ including position and momentum of the oscillator and $u^*(t)$ denotes the trajectory that consists of the same states visited in reverse order with inverted momenta.
Because no work is done on the system, the heat $\Q$ exchanged along a trajectory equals the energy lost by the system, $\Q = -[E(t)-E(0)]$, where $E(0)$ and $E(t)$ are the energy at the beginning and at the end of the stochastic trajectory.

The fluctuation theorem holds for any time $t$ at which $\dS $ is evaluated, and it is not required that the system has reached the equilibrium distribution at time $t$.
In general, the steady distribution $P_0(u(t))$ necessary to compute $\Delta\entropy$ is unknown. However, from the distribution derived for our model Eqn.~\eqref{eq:Heff}, we find that for relaxation from a non-equilibrium steady state generated by nonlinear feedback of strength $\fb$ and parametric modulation of strength $\emod_0$, the relative entropy change is given by \cite{Gieseler:2015bl, GieselerThesis}
\begin{equation}\label{eq:dS}
  \dS=
 \beta \Delta \Efb
\end{equation}
where $\Delta \Efb = \Efb(t) -\Efb(0) $ and $\Efb = \Heff-E$ is the contribution to the effective energy \eqref{eq:Heff} from the modulation of the trapping laser.
Thus, our stochastic model (c.f. Eqn.~\eqref{eq:dS}) allows us to express the relative entropy change during a relaxation trajectory in terms of the energy at the beginning and the end of that trajectory.
Note that $\Efb$ can also be negative, e.g. when we cool the particle motion.
For example, setting $\phimod = \pi/4, \eta = 0$ in Eqn.~\eqref{eq:Heff} amounts to cooling with phase looked loop feedback \cite{Jain2016a}. In this case, $\Efb(t) = \left(\emod\wo \left/2\g{CM}\right.\right) E(t) = (\beta_\text{eff}-\beta) E(t)/\beta$ and consequently $\Delta \mathcal{S} = (\beta-\beta_\text{eff})Q$. Since here the relative entropy change is proportional to the heat $\Q$, it is intuitively clear that $\dS$ is a measure for the dissipation during the relaxation process.
This dissipation function was measured in the underdamped regime with optical tweezers and colloidal particles and was one of the very first experimental demonstrations of the validity of a fluctuation theorem \cite{Wang:2002hw}.
In contrast, for parametric feedback, where $\emod = 0$, $\eta > 0$, we find that $\dS\propto E^2(t)-E^2(0)$ is no longer proportional to $\Q$.
In this case, $\dS$ still is a measure for the reversibility of the relaxation process. However, it is no longer a simple function of the exchanged heat.
This was verified in the underdamped regime using a~levitated nanoparticle by Gieseler {et al.,} \cite{Gieseler2014}, when starting from a variety of non-equilibrium steady states (c.f. Fig.~\ref{fig:FluctuationTheorem}). 
It is important to note that the relative entropy change defined here is a special case of an infinite class of quantities $R$, which have been introduced in their general form in a classic paper by Seifert \cite{Seifert:2005fu} and has been called the dissipation function $\Omega_t$ by Evans and Searles \cite{Evans:2002tg}. Because it depends on the stochastic trajectory it is commonly referred to as ``stochastic entropy''.

Later Hoang et al., \cite{Hoang:2018fn} experimentally demonstrated another differential fluctuation theorem with levitated nanoparticles
\begin{equation}
  \frac{P\left(-W, u^*(t)\right)}{P\left(W, u(t)\right)} 
  = e^{-\beta(W-\Delta F)},
\end{equation}
which determines the probabilities that the system performs work $W = -\int_0^\tau   \dot{f}(t) q(t)   \d t$ against an~external force $f(t)$.
In the experiment, the force is ramped from $f_\text{off}$ to $f_\text{on}$ at a rate that is much faster than the velocity and position relaxation times, such that when the ramp finishes, the system is far from thermal equilibrium.
The free energy difference between the equilibrium states at the beginning and at the end of the ramp is given by $\Delta F = -(f_\text{on}^2-f_\text{off}^2) / (2m\Omega^2)$.

Note that the differential fluctuation theorems can be integrated to yield integral fluctuation theorems $\langle\exp(-R)\rangle=1$, such as the Jarzynski equality \cite{Jarzynski1997, Liphardt:2002uia} and its refined version, the~Hummer-Szabo relation \cite{Hummer:2001ts, Gupta:2011cp}, which allows the reconstruction of free energy potentials.
Thus, by verifying the underlying differential fluctuation theorem, the validity of the integral fluctuation theorem is implied and consequently also the ``second law'' inequality  $\langle R\rangle\geq 1$~\cite{SeifertReview}. Importantly, the fluctuation theorems are valid for arbitrarily-far-from-equilibrium processes. Both~detailed and integral fluctuation theorems allow the estimation of equilibrium free energy changes from nonequilibrium protocols and have found applications in determining the free energies of DNA molecules \cite{Liphardt:2002uia, Collin:2005fx}.
For a detailed review see Refs.~\cite{SeifertReview, MartinezReview, CilibertoReview, RitortReview}.

\section{Heat Engines\label{sec:HeatEngines}}

In the previous section we saw that in microscopic systems, thermodynamic quantities such as the work against an external force and the heat exchanged with the environment, become stochastic quantities due to the underlying fluctuating trajectories through phase space. Yet work, heat and efficiency can be rigorously defined within the framework of stochastic thermodynamics, yielding the respective ensemble quantities after averaging \cite{SeifertReview, Sekimoto:2010uua}.
As a consequence, the output of a microscopic engine will be fluctuating with the possibility of it running ``{{\emph{in reverse}}}''.
Interestingly, one can draw analogies between a particle in an optical trap and an ideal gas inside a piston, where the trap stiffness is analogous to the inverse of an effective volume while the variance of the trajectory of the particle can be seen as an effective pressure.
By monitoring the motion of a particle as it undergoes the cyclic heat engine, one can extract the work statistics.
This is the basic idea behind the following stochastic heat engines.
We leave a full discussion of the heat and entropy statistics to other sources, for example Spinney \& Ford \cite{FordBook}.

\begin{figure}[h]
\begin{center}
	\includegraphics[width=\textwidth]{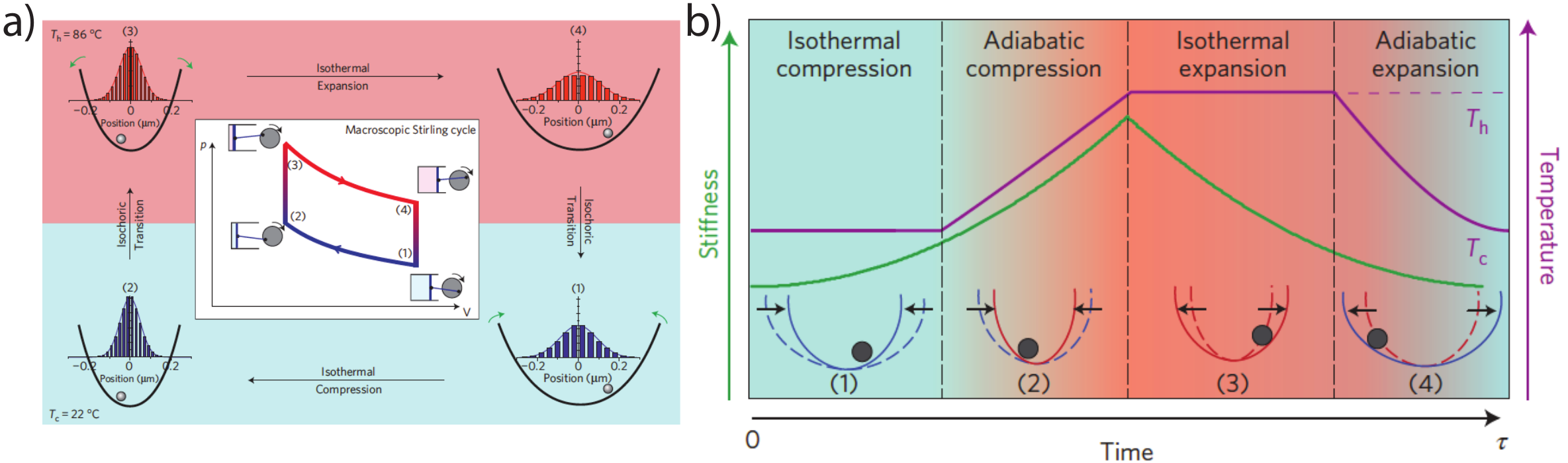}
  \caption{{{Single particle engines}}.
  (\textbf{a}) Realization of a Stirling engine by Blickle \& Bechinger \cite{Blickle2011} using laser absorption to change the temperature of the environment. 
  (\textbf{b}) Martinez et al., \cite{Martinez:2015gq} realized a~microscopic Carnot engine. The adiabatic steps of the Carnot engine requires to change the temperature and the trap stiffness synchronously.
  Figures reproduced from Nature Physics.
   \label{fig:Carnot}}
\end{center}
\end{figure}

Schmiedl \& Siefert gave the first full description of a colloidal stochastic heat engine \cite{Schmiedl2008} and the first experimental realization was by Blickle \& Bechinger \cite{Blickle2011}, who locally heated the liquid (water) surrounding an optically trapped particle through laser absorption (c.f. Fig.~\ref{fig:Carnot}a). Thereby, they~realized temperature changes of 70 $^\circ$C in 10\,ms, and their data agreed well with theoretical predictions. This study was of particular importance, since it clearly demonstrated the fluctuating nature of the work statistics and observed that for \emph{{individual}} cycles sometimes the engine operated in reverse, due~to the fluctuating position statistics, thereby demonstrating that microscopic heat engines behave fundamentally differently from their macroscopic counterparts.
The fluctuations of the power can be accounted for in a power fluctuation theorem \cite{MartinezReview} and a stochastic definition of efficiency is given by the ratio of the stochastic work extracted in a cycle and the stochastic heat transferred from the hot bath to the system.

The efficiency of the Stirling engine is fundamentally limited by the isochoric steps, where heat is transferred between the system and the heat baths that are at different temperatures, making the cycle inherently irreversible.
The Carnot engine overcomes this limit by replacing the isochoric steps with adiabatic changes, during which no heat is exchanged with the environment.
The second law of thermodynamics imposes a maximum (Carnot) efficiency $\eta_C = 1-\T{C}/\T{H}$ that can be reached by any heat engine operating between two baths at temperature $\T{C}$ and $\T{H}$, respectively \cite{Carnot1978}.

However, it is commonly believed that the realization of an adiabatic change requires that the control to the system is applied extremely slowly and therefore in a Markovian process the Carnot efficiency can only be achieved in the limit of zero power output \cite{Shiraishi2016}.
For practical applications, however, one is interested in the efficiency at maximum power, which has lead to the birth of finite-time thermodynamics \cite{Andresen:1984ks}.
For the ideal case, the efficiency at maximum power is limited by Novikov-Curzon-Ahlborn efficiency \cite{Novikov1958,Curzon1975} $\eta^* = 1-\sqrt{\T{C}/\T{H}}$, which is smaller than the Carnot efficiency $\eta^* < \eta_C$.
Later it was found that it is possible to generate a shortcut to adiabaticity \cite{Berry:2009gi, Demirplak:2003ga}.
In these protocols the evolution of the system mimics the adiabatic dynamics without the requirement of slow driving by introducing a counterdiabatic driving term, raising the question whether the Novikov-Curzon-Ahlborn efficiency can be surpassed \cite{Schmiedl2008}. 
As a consequence, optimal protocals that lead to shortcuts to adiabaticity have received much attention recently \cite{Deffner:2014iv, Jarzynski:2013do, delCampo:2013jp, Tu:2014it}, both for their experimental relevance and as an interesting theoretical problem in its own right.

Recent studies have also asked whether single parameter bounds such as the Novikov-Curzon-Ahlborn bound is the best metric, discussing instead “trade-offs”  between efficiency and power under different experimental conditions and under non-equilibrium operation  \cite{Brandner2015}.
Levitated nanoparticles could test these relations over a large parameter space. 
For instance, one can explore non-Markovian dynamics \cite{Rosinberg2015} due to feedback and operation under periodic temperature variations \cite{Brandner2015} which could be achieved through modulation of the laser beam as we discussed before.
Besides, the full over- to under-damped regime is easily accessible whilst dynamically varying all of the relevant thermodynamics quantities (such as trapping volume and temperature).
In addition, the tantalizing potential to operate in the quantum regime could enable exploration of constraints on the efficiency and power production of non-Markovian quantum engines \cite{Shiraishi2017}.

To overcome the limitations of the Stirling cycle, Martinez et al. implemented a Carnot cycle with an optically trapped colloidal particle (c.f. Fig.~\ref{fig:Carnot}b). The adiabatic ramp was thereby realized by changing the temperature and trap stiffness together such that the ratio $T^2/\ks{}$ remained \mbox{constant \cite{Martinez2015,Tu:2014it}}.
This protocol required a precise control over the bath temperature that is synchronized with the change of the trap stiffness. This is not possible with heating of the surrounding water as done by Blickle \& Bechinger \cite{Blickle2011}. Instead they produced an effective hot temperature bath with fluctuating electromagnetic fields as we discussed earlier \cite{Martinez:2013fe}.
For slow driving, their Carnot engine attained the fundamental limit given by the Carnot efficiency and the efficiency at maximum power was in excellent agreement with the Novikov-Curzon-Ahlborn efficiency.
In addition, they showed that the Carnot bound can be surpassed for a small number of non-equilibrium cycles \cite{Martinez:2015gq}.
\mbox{For a detailed} discussion of microscopic heat engines in the overdamped regime we refer to Refs.~\cite{Dinis:2016gx, CilibertoReview, MartinezReview}.

The implementations discussed so far have been realized with colloidal systems, where the motion of the particle is heavily damped due to the close contact with the surrounding liquid.
Under these conditions, a measurement of the momentum distribution is very challenging, although not impossible \cite{Huang:2011bb, Kheifets:2014hq}.
Martinez et al., \cite{Martinez2015} circumvented this challenge by extrapolating the instantaneous velocity from the mean-squared time-averaged velocity.
However, from a fundamental standpoint it is desirable to have direct access to the instantaneous position {and} momentum of the particle.
%
The instantaneous momentum of the particle can be measured easily by operating under high vacuum conditions  \cite{Li2010, Gieseler:2012bi}.
In addition, the investigation of much more isolated systems provides a path toward the future realization of quantum heat engines and it has been suggested that super Carnot efficiencies can be attained by clever bath engineering \cite{Rossnagel:2014di}.

From our previous discussion it is clear that optically levitated particles are well suited to implement a microscopic heat engine in the underdamped regime.
Analogous to the experiments with colloidal particles \cite{Martinez:2015gq, Blickle2011}, the volume of the engine is controlled through the power of the trapping laser (c.f. Eqn.~\eqref{eqn:frequencies_translational}).
However, due to the weak interaction with the environment, the effective temperature can be controlled with much higher precision (c.f. Eqn.~\eqref{eq:Teff}) through a combination of gas pressure, external driving and feedback cooling and even allows to create non-thermal \mbox{baths \cite{Gieseler2014, Gieseler:2015bl, Rashid:2016hp}} that could lead to surpassing the Carnot efficiency \cite{Rossnagel:2014di}.
Such an~all-optical heat engine was proposed by Dechant {et al.}, \cite{Dechant2014}. In this proposal, cooling is realized through the interaction with an optical cavity instead of active feedback cooling.
The all-optical control provides flexibility in optimizing the heat engine for maximizing its performance.
In the overdamped case the analytic treatment of optimal protocols is possible because the dynamics can be described by a simplified in terms of the slow position variable \cite{Schmiedl2008}.
In contrast, this is not possible in the underdamped case, where the position and velocity variables cannot be separated \cite{GomezMarin2008, Dechant2017}, and~numerical methods must be used.
An important note is that these optimal protocols increase the power output and the efficiency of the engine by introducing rapid changes in the trapping frequency. Being able to realize this kind of fast control experimentally is a distinct advantage of the all-optical nature of the heat engine. In particular, a heat engine realized with a fast cavity response or a cavity-free setup could prove advantageous, since the control is not limited by the finite cavity response time.

\section{Conclusions}

In this review we have explored the potential for levitated nanoparticles to address questions in stochastic thermodynamics and non-equilibrium physics on the single particle level in the underdamped regime.

In the overdamped regime, micron-sized colloidal particles in liquid have already been used extensively to construct micro-engines and to study the statistical properties of their power and efficiency, largely motivated by trying to understand biological systems.
A better understanding of how nature builds machines and motors at the molecular level then allows scientists to build their own molecular devices \cite{Browne2006, ErbasCakmak2015}.
Although these objects operate in overdamped environments, the~timescale of their operation requires one to account for inertial contributions \cite{GomezMarin2008}. As~discussed, measuring the instantaneous momentum in real time is not straightforward in overdamped systems.
In~contrast, the momentum relaxation in underdamped systems is much slower due to the weak interaction with the environment, which allows for obtaining a~complete picture of the dynamics.
Hence,~levitated particles may offer new insights into the molecular~world.

In levitated systems, deterministic forces and stochastic forces are well controlled experimentally, thus giving access to new parameter regimes.
For example, levitated nanoparticles have led to the first observation of ballistic Brownian motion \cite{Li2010} and to the first quantitative measurement of Kramers turnover \cite{Rondin2017} in addition to demonstrating general fluctuation theorems in the underdamped regime~\cite{Gieseler2014, Hoang:2018fn}.

The underdamped regime is of fundamental interest, since the underlying equations of motion contain inertia which in overdamped systems is typically ignored.
It also allows one to make the connection to the even more fundamental unitary evolution of quantum mechanical systems.
Therefore, future experiments with levitated nanoparticles will help to characterize the sources of irreversibility in micro-engines and give new insight into the statistical properties of their efficiencies that could inspire new strategies in the design of efficient nano-motors.
In addition, rapid progress in cooling the center-of-mass motion will enable the operation in the quantum regime, thereby realizing a~textbook quantum Brownian particle.

In the quantum regime, the information we can extract from a system is limited by the Heisenberg uncertainty principle \cite{Heisenberg:1927vy}. The fact that information is physical is also well established in thermodynamics through Landauer's principle, which asserts that there is a minimum possible amount of energy required to erase one bit of information \cite{Landauer:1961}.
This link between information theory and thermodynamics was verified experimentally with a colloidal system where the information was obtained by light scattering from the particle  \cite{Lutz:2015ke, Berut2012}.
In ultra-high vacuum, the interaction with the optical light field is the dominant interaction of the levitated particle with its environment~\cite{Jain2016a}.
Therefore, the quantum back-action of the measurement \cite{Braginsky1992, Purdy:2013cb} starts to play a role in this regime.
The impact of the measurement process in the operation of heat engines and work extraction \cite{Talkner:2016bj} is a very active field of research that combines information theory, the quantum measurement process and thermodynamics.
Levitated particles in high vacuum are already exploring thermodynamics in the underdamped regime and are poised for venturing into the quantum regime. 
However, definite theoretical proposals for the realization of thermodynamic protocols with levitated nanoparticles in the quantum regime are still lacking. We hope that this review changes this by making the fundamentals of levitated nanoparticles easily accessible to the community of quantum thermodynamics while also raising awareness of this exciting field among researchers working in levitation.

\end{document}

%% file: commands.tex
\newcommand{\mum}{\mu \text{m}} 
\newcommand{\mbar}{\text{mbar}} 
\newcommand{\Hz}{\text{Hz}} 

\renewcommand{\d}{{\rm d}}
\newcommand{\kB}{{\rm k_B}}
\newcommand{\waist}{\text{w}}
\newcommand{\w}{\Omega}
\newcommand{\wo}{\Omega_0}

\newcommand{\Pg}{P_{\rm gas}}
\newcommand{\cg}{{\rm c_{\rm P}}}
\newcommand{\crosssectiongas}{\sigma_\text{gas}}
\newcommand{\viscosity}{\mu_v}
\newcommand{\mgas}{m_\text{gas}}
\newcommand{\eo}{\epsilon_0}
\newcommand{\er}{\epsilon_{\rm r}}
\newcommand{\acccoeff}{c_{\rm acc}}
\newcommand{\whn}{\Xi} 

\newcommand{\g}[1]{\Gamma_{\rm #1}}
\newcommand{\T}[1]{T_{{\rm #1}}}

\newcommand{\dpcoeff}{c_{\rm dp}}
\newcommand{\Pscat}{P_{\rm scat}}
\newcommand{\Popt}{P_{\rm opt}}
\newcommand{\Poptm}{\bar{P}_{\rm opt}}
\newcommand{\Iopt}{I_{\rm opt}}

\newcommand{\wopt}{\omega_{\rm L}}
\newcommand{\kopt}{k_{\rm L}}
\newcommand{\lambdaopt}{\lambda_{\rm L}}
\newcommand{\polar}[1]{\alpha_{\rm #1}}
\newcommand{\scatcross}{\sigma_{\rm scat}}
\newcommand{\totcross}{\sigma_{\rm tot}}
\newcommand{\F}[1]{\mathbf{F}_{\rm #1}}

\newcommand{\ks}[1]{{\rm k_{\rm #1}}} 

\newcommand{\Qf}{{\rm Q}} 

\newcommand{\Sff}[1]{S_{\rm ff}^{#1}}
\newcommand{\noise}{n}

\newcommand{\chitensor}{\boldsymbol{\chi}}

\newcommand{\depol}{\boldsymbol{N}}

\newcommand{\q}{q}

\newcommand{\var}[1]{\sigma_{#1}^2}
\newcommand{\corr}[2]{\langle #1(t)#2(0) \rangle}

\renewcommand{\a}{A} 

\newcommand{\R}[1]{{\rm R}_{#1}} 

\newcommand{\emod}{\zeta} 
\newcommand{\phimod}{\phi_{\rm mod}} 
\newcommand{\omegamod}{\Omega_{\rm mod}} 
\newcommand{\fb}{\eta} 

\newcommand{\dS}{\Delta\mathcal{S}} 
\newcommand{\Q}{\mathcal{Q}} 
\newcommand{\entropy}{\Phi} 

\newcommand{\diag}{\text{diag}}

\newcommand{\Heff}{H_\text{eff}}
\newcommand{\Efb}{H_\text{fb}}